\pdfoutput=1
\documentclass[fleqn,usenatbib]{mnras}

\usepackage{newtxtext,newtxmath}

\usepackage[T1]{fontenc}

\usepackage[dvipsnames]{xcolor}
\usepackage[normalem]{ulem}
\usepackage{bm}
\usepackage[utf8]{inputenc}
\usepackage{pgfplots}
\usepackage{tikz}
\usepgfplotslibrary{groupplots}
\pgfplotsset{compat=1.14}
\usepackage{capt-of}
\usepackage{graphicx}

\usepackage{amssymb}

\usepackage{mathrsfs}
\usepackage{amsmath}
\usepackage{hyperref}
\usepackage{lineno}
\usepackage{subfigure}
\usepackage{widetext}

\definecolor{pink}{RGB}{255, 20, 147}      

\usepackage{subfigure}

\newcommand{\Lagr}{\mathcal{L}}
\definecolor{comment}{RGB}{166, 38, 164}

\definecolor{pink}{RGB}{255, 20, 147}

\definecolor{carminered}{rgb}{1.0, 0.0, 0.22}

\definecolor{byzantine}{rgb}{0.74, 0.2, 0.64}

\definecolor{blue-violet}{rgb}{0.30, 0.1, 0.89}

\definecolor{amethyst}{rgb}{0.6, 0.4, 0.8}

\definecolor{blue(munsell)}{rgb}{0.0, 0.5, 0.69}

\definecolor{blue-violet}{rgb}{0.54, 0.17, 0.89}

\title[Tidally Resonant Oceans: GRB 211211A]{Gamma-ray burst precursors from tidally resonant neutron star oceans: potential implications for GRB 211211A}

\author[A.G. Sullivan et al.]{
Andrew G. Sullivan,$^{1}$\thanks{E-mail: ags2198@stanford.edu} 
Lucas M. B. Alves,$^{2}$
Zsuzsa M\'arka,$^{3}$
Imre Bartos,$^{4}$
\newauthor
and Szabolcs M\'arka$^2$ 
\\
$^{1}$Kavli Institute for Particle Astrophysics and Cosmology, Department of Physics, Stanford University, Stanford, CA 94305, USA\\
$^{2}$Department of Physics, Columbia University in the City of New York, New York, NY 10027, USA\\
$^{3}$Columbia Astrophysics Laboratory, Columbia University in the City of New York, New York, NY 10027, USA\\
$^4$Department of Physics, University of Florida, Gainesville, FL 32611, USA
}

\date{Accepted XXX. Received XXX; in original form XXX}

\pubyear{2023}

\begin{document}
\label{firstpage}
\pagerange{\pageref{firstpage}--\pageref{lastpage}}
\maketitle

\begin{abstract}
Precursors have been observed seconds to minutes before some short gamma-ray bursts. While the precursor origins remain unknown, one explanation relies on the resonance of neutron star pulsational modes with the tidal forces during the inspiral phase of a compact binary merger. In this paper, we present a model for short gamma-ray burst precursors which relies on tidally resonant neutron star oceans. In this scenario, the onset of tidal resonance in the crust-ocean interface mode ignites the precursor flare, possibly through the interaction between the excited neutron star ocean and the surface magnetic fields. From just the precursor total energy, the time before the main event, and a detected quasi-periodic oscillation frequency, we may constrain the binary parameters and neutron star ocean properties. Our model can immediately distinguish neutron star-black hole mergers from binary neutron star mergers without gravitational wave detection. We apply our model to GRB 211211A, the recently detected long duration short gamma-ray burst with a quasi-periodic precursor, and explore the parameters of this system. The precursor of GRB 211211A is consistent with a tidally resonant neutron star ocean explanation that requires an extreme-mass ratio neutron star-black hole merger and a high mass neutron star. While difficult to reconcile with the main gamma-ray burst and associated kilonova, our results constrain the possible precursor mechanisms in this system. A systematic study of short gamma-ray burst precursors with the model presented here can test precursor origin and probe the possible connection between gamma-ray bursts and neutron star-black hole mergers.
\end{abstract}

\section{Introduction}
\label{sec:Introduction} %
Short gamma-ray bursts (sGRB) represent electromagnetic counterparts to compact binary mergers \citep{1992ApJ...395L..83N, 2017ApJ...848L..12A, 2017ApJ...848L..13A}. In these events, the powerful dynamics in the binaries can form relativistic jets which produce gamma-ray emission. While the exact gamma-ray burst (GRB) central engine$-$whether a strongly magnetized proto-neutron star \citep[e.g.]{2020MNRAS.495L..66C, 2021MNRAS.502.2482S} or a black hole \citep{2021GReGr..53...59S}$-$is unknown, compact binary mergers should produce sGRBs if they contain a neutron star and relativistic jets form at the time of the merger \citep{2022PhRvD.105h3004S}. {Some}
sGRBs are followed by {kilonovae} \citep[e.g.]{2020MNRAS.493.3379R}, optical thermal emission from the radioactive decay of heavy elements produced by the merger \citep{1998ApJ...507L..59L, 2010MNRAS.406.2650M, 2012ApJ...746...48M, 2013ApJ...774...25K, 2013Natur.500..547T, 2018IJMPD..2742004C}. With the additional detection of gravitational waves (GWs) \citep{1994PhRvD..49.2658C,2017PhRvL.119p1101A} and possibly high energy neutrinos \citep{2003MNRAS.342..673R, 2021arXiv211113005C, 2023ApJ...944...80A}, multimessenger observations of compact binary mergers reveal the properties of their progenitors as well as the dynamical processes at work during the inspirals.

Precursor electromagnetic emission has been associated with some observed GRBs, including $\gtrsim1\%$ of sGRBs \citep{2010ApJ...723.1711T, 2019ApJ...884...25Z, 2020PhRvD.102j3014C, 2020ApJ...902L..42W, 2021ApJS..252...16L}. Such precursors can be $\sim1-100$ s prior to the main sGRB event \citep{2010ApJ...723.1711T} and may indicate particular features of the systems from which they originate. Proposed mechanisms for producing sGRB precursors include an initial episode of the main GRB emission \citep{2015MNRAS.448.2624C}, the interaction between neutron star magnetospheres \citep{2021JPlPh..87a8402A}, the orbital motion of a weakly magnetized companion and a highly magnetized neutron star  \citep{1996ApJ...471L..95V,  2001MNRAS.322..695H, 2011ApJ...742...90M, 2012ApJ...757L...3L, 2012ApJ...755...80P, 2021MNRAS.501.3184S}, and the tidally induced shattering of a neutron star crust \citep{2012PhRvL.108a1102T, 2020PhRvD.101h3002S, 2020PhRvD.101j3025G, 2021MNRAS.504.1273P, 2021MNRAS.506.2985K, 2021MNRAS.508.1732K, 2022MNRAS.514.5385N}. Particularly, resonant tides, in which the binary orbital motion becomes resonant with internal neutron star pulsational modes, represent promising causes of crustal shattering and by extension electromagnetic emission \citep{2012PhRvL.108a1102T, 2021MNRAS.504.1273P, 2022MNRAS.514.5385N, 2023arXiv230702996D}. 

Lower frequency modes in the surface layers of a neutron star have the potential to generate early precursors when resonant with orbital motion. A possible site of early resonance is the fluid outer layer, the neutron star's ocean. The ocean sustains its own set of low frequency pulsational modes associated with its lower density and separation from the rest of the neutron star by the elastic crust \citep{1995ApJ...449..800B, 2001ApJ...550..426L, 2005ApJ...619.1054P, 10.1093/mnras/stad389}. Tidal resonance should occur in neutron star ocean modes early in compact binary inspirals and may deposit large amounts of energy into the modes \citep{10.1093/mnras/stad389}. In fact,
\cite{10.1093/mnras/stad389} show that the energy deposited into the neutron star ocean during tidal resonance may be sufficient to produce a detectable electromagnetic flare. 

Motivated by the results of \cite{10.1093/mnras/stad389}, we advance a new model for sGRB precursors in this paper. We propose that the interface pulsational mode associated with the crust-ocean boundary may become resonant during the inspiral phase of a compact binary merger. The excited ocean consequently represents the site of a sGRB precursor. This model can be applied to sGRBs from both binary neutron star (BNS) mergers and neutron star-black hole (NSBH) mergers, thus admitting precursors from either scenario and distinguishing between them through electromagnetic emission alone.

In this paper, we develop analytical formulae for precursor sGRB observables which can be used to estimate compact binary parameters in the context of this model. As a first application, we consider the precursor associated with the recently detected GRB 211211A, an especially unique sGRB due to its long length \citep{2022Natur.612..223R} as well as the {possible} identification of quasi-periodic oscillations (QPOs) in its precursor emission \citep{2022arXiv220502186X}. In Section \ref{sec:tidalresNSO}, we review tidal resonance in neutron star oceans, and present the theory relevant to our model. In Section \ref{sec:ModelSummary}, we provide a detailed discussion of our precursor model. In Section \ref{sec:GRB211211A}, we apply our model to GRB 211211A and constrain the parameters of the system. We also evaluate our model's applicability to this system and its potential consequences. In Section \ref{sec:Conclusion}, we consider future prospects and conclude.

\section{Tidal Resonances in Neutron Star Oceans}
\label{sec:tidalresNSO} 
In this section, we review the general properties of neutron star ocean tidal resonances \citep{10.1093/mnras/stad389}. Neutron stars, like main sequence stars, possess a spectrum of excitable pulsational modes \citep{1988ApJ...325..725M, 1994ApJ...426..688R, 1994MNRAS.270..611L, 2006PhRvD..73h4010P, 2007CQGra..24.4147S, 2012MNRAS.419..638P}. A select few modes are localized almost entirely to the neutron star ocean, including surface $g$-modes \citep{1995ApJ...449..800B}, as well as the crust-ocean interface mode or $i$-mode \citep{2005ApJ...619.1054P, 10.1093/mnras/stad389}. The $i$-mode is the generalization of the shallow ocean surface mode to the case where the ocean floor is not completely rigid. Thus, it is associated with the discontinuity in shear modulus across the crust-ocean boundary. Tides in compact binary systems represent a promising mechanism for exciting these modes \citep{1994MNRAS.270..611L, 2012PhRvL.108a1102T, 2013ApJ...777..103T, 2021MNRAS.504.1273P, 10.1093/mnras/stad389}. Tidal resonance, which occurs in compact binary inspirals when a harmonic of the orbital frequency matches the mode frequency, can deposit significant energy into the crust-ocean $i$-mode, potentially sufficient to produce a flare \citep{10.1093/mnras/stad389}. 

\subsection{Neutron Star Ocean Modes}
Quantitatively, the modes are fluid perturbations on the background neutron star. The principle equation of motion is the perturbed Euler equation
\begin{equation}
\label{eq:euler}
    \partial_t^2 \vec{\xi} + \frac{\nabla \delta p}{\rho}-\frac{\delta \rho}{\rho^2}\nabla p-\frac{1}{\rho} \nabla \cdot \boldsymbol{\sigma}=-\nabla \chi,
\end{equation}
where $\vec{\xi}$ is the Lagrangian fluid displacement, $\rho$ is the background fluid density, $p$ is the background fluid pressure, $\delta \rho$ is the Eulerian perturbation of the density, $\delta p$ is the Eulerian perturbation of the pressure, $\boldsymbol{\sigma}=\sigma_{ij}$ is the elastic stress tensor, and $\chi$ is an external potential, which corresponds to the tidal potential in this case. The continuity equation combined with the definition of the Lagrangian perturbation gives an additional governing equation \citep{1978ApJ...221..937F}
\begin{equation}
\label{eq:density2}
    \Delta \rho=\delta \rho + \vec{\xi}\cdot \nabla \rho =-\rho \nabla \cdot \vec{\xi}.
\end{equation}
Setting $\chi=0$, the mode solutions have $\vec{\xi}\propto e^{i \omega t}$, where $\omega$ is the mode frequency. Eq. \ref{eq:euler} simplifies to the eigenvalue equation
\begin{equation}
\label{eq:eigen}
    (\Lagr-\omega^2 \rho)\vec{\xi}=0,
\end{equation}
where $\Lagr$ is the operator which contains the non-time derivative terms in eq. \ref{eq:euler}. 
\subsubsection{Mode Frequency}
The shallow ocean surface mode corresponds to an approximate analytical solution to eqs. \ref{eq:euler} and \ref{eq:density2}. The shear modulus is $\Tilde{\mu}=0$ in the fluid ocean, so eq. \ref{eq:euler} reduces to 
\begin{equation}
    \partial_t^2 \vec{\xi} + \frac{\nabla \delta p}{\rho}-\frac{\delta \rho}{\rho^2}\nabla p=0.
\end{equation}
At the surface of the star, the traction vanishes, so
\begin{equation}
    \Delta p=\delta p +\vec{\xi}\cdot \nabla p=0
\end{equation}from the definition of the Lagrangian displacement. Therefore, in a shallow ocean
\begin{equation}
    \delta p\approx-\vec{\xi}\cdot \nabla p=\xi_r\rho g, 
\end{equation}
where $\xi_r$ is radial component of $\vec{\xi}$ and $g=\frac{G M_\star}{R_\star^2}$ is the magnitude of the gravitational acceleration at the surface. The second equality arises from assuming the background ocean is in hydrostatic equilibrium.
In a shallow ocean, we expect $\delta \rho/\rho<<1$
\citep{2023arXiv230405413G}. The Euler equation consequently simplifies to 
\begin{equation}
\label{eq:simplifiedEuler}
      \partial_t^2 \vec{\xi} + g\nabla  \xi_r=0,
\end{equation}
while eq. \ref{eq:density2} simplifies to
\begin{equation}
    \frac{d\rho}{dr} \xi_r=-\rho \nabla \cdot \vec{\xi},
\end{equation}
where we have assumed that the background $\rho$ is purely radial. In the ocean, we expect $\frac{d \rho}{dr}\approx\frac{\rho}{h_o}$, where $h_o$ is the ocean depth, so we obtain an expression for $\xi_r$ 
\begin{equation}
    \xi_r=-h_o \nabla \cdot \vec{\xi}.
\end{equation}
Substituting this value for $\xi_r$ into eq. \ref{eq:simplifiedEuler} gives 
\begin{equation}
      \partial_t^2 \vec{\xi} -gh_o\nabla  (\nabla\cdot\vec{\xi})=0.
\end{equation}
Assuming $\xi$ is curl free by construction gives a wave equation 
\begin{equation}
      \partial_t^2 \vec{\xi} -gh_o\nabla^2  \vec{\xi}=0.
\end{equation}
Restricting to the value of $\xi$ at the ocean surface and expanding in spherical harmonics yields the ordinary differential equation
\begin{equation}
    \partial_t^2 \vec{\xi}+\frac{l(l+1)g h_o}{R_\star^2}\vec{\xi}=0,
\end{equation}
whose solution is a simple harmonic oscillator with frequency \begin{equation}
\label{eq:frequency}
    \omega=\sqrt{l(l+1)\frac{G M_\star}{R_\star^3}\frac{h_o}{R_\star}}.
\end{equation}
Therefore, the surface fluid layer of the neutron star naturally sustains periodic oscillations.

In this analysis, we have assumed that the crust is completely rigid at the boundary between the neutron star ocean and crust. In general, the crust should be elastic with a non-infinite shear modulus $\breve{\mu}$ \citep{2009PhRvL.102s1102H, 2011MNRAS.416...22B, 2023MNRAS.518.3813Z}. \cite{2005ApJ...619.1054P} showed that when the neutron star crust's shear modulus is less than the pressure at the crust-ocean boundary $p_o$, the shallow ocean frequency given by eq. \ref{eq:frequency} must be corrected by a factor of $\sqrt{\breve{\mu}/p_o}$. Therefore, the mode frequency is
\begin{equation}
\label{eq:frequencymod}
    \omega=\sqrt{\frac{\breve{\mu}}{p_o}l(l+1)\frac{G M_\star}{R_\star^3}\frac{h_o}{R_\star}}.
\end{equation}
Evidently, the shallow ocean surface mode will pulsate with frequency given by eq. \ref{eq:frequencymod}, which depends on the parameters of the neutron star. 

\subsection{Tidal Resonance}
The tidal potential $\chi$ induced by a companion object is
\citep{1977ApJ...213..183P, 1994MNRAS.270..611L}
\begin{equation}
\label{eq:tidal}
    \chi=-\sum_{l=2}^{\infty}\sum_{m=-l}^l \frac{G M r^l}{D(t)^{l+1}}W_{l m} e^{-i m \Phi(t)} Y_{l m}(\theta, \phi), 
\end{equation}
where $M$ is the companion mass, $D(t)$ is the binary separation, $\Phi(t)$ is the true anomaly, and $W_{l m}$ is the numerical coefficient \citep{1977ApJ...213..183P}
\begin{equation}
    W_{l m}=(-1)^{\frac{l+m}{2}}\frac{\left((\frac{4 \pi}{2l+1})(l-m)!(l+m)!\right)^{\frac{1}{2}}}{2^l (\frac{l-m}{2})!(\frac{l+m}{2})!},
\end{equation}
where $l+m$ must be even. When this potential is added, the Euler equation can be expressed as 
\begin{equation}
\label{eq:ordinarydiffeq}
    (\Lagr+\rho\partial_t^2 )\vec{\xi}=-\rho\nabla \chi,
\end{equation}
where $\Lagr$ is defined by eq. \ref{eq:eigen}.
We assume this equation has the solution
\begin{equation}
    \vec{\xi}=\sum_n a_n(t) \vec{\xi}_n,
\end{equation}
where $\vec{\xi}_n$ is the eigenvector solution to eq. \ref{eq:eigen}. From eq. \ref{eq:ordinarydiffeq} and the orthogonality condition  $\int\rho\vec{\xi}_n^*\cdot\vec{\xi}_m dV=A_{n}^2 \delta_{nm}$ \citep{10.1093/mnras/stad389}, we obtain an equation for $a_n(t)$
\begin{equation}
\label{eq:ODEa}
    \ddot{a}_n(t)+\omega_n^2 a_n(t)=\frac{G M W_{lm}}{D(t)^{l+1} }e^{-im \Phi(t)} \frac{Q_{nl}}{A_n^2}.
\end{equation}
where $Q_{nl}$ is the overlap integral defined by 
\begin{equation}
    Q_{nl} = \int \rho \vec{\xi}_n^* \cdot\nabla(r^l Y_{lm}(\theta, \phi)) d V .
\end{equation}
The mode most likely to be tidally excited is the $l=2$ mode because the driving tidal force is lowest order in $1/D$. 

\subsubsection{Resonance Time}
As is typical for driven harmonic oscillators, a resonant oscillation will occur when $m\Dot{\Phi}(t)=\omega_n$. In the case of an inspiraling compact binary, $\Dot{\Phi}(t)$ continuously increases, so the orbital tidal force will become resonant with the $i$-mode at some point before the merger. The time before merger at which resonance occurs can be computed by recalling 
\begin{equation}
    \label{eq:orbitalfrequencyD}
    \dot{\Phi}(t)=\sqrt{\frac{G(M+M_\star)}{D(t)^3}},
\end{equation}
for circular binaries. The orbital separation at the time of resonance is
\begin{equation}
    D_r=\left(\frac{m^2 G(M+M_\star)}{\omega_n^2}\right)^{\frac{1}{3}}.
\end{equation}
The time until merger due to gravitational wave emission for a given orbital separation $D$ is \citep{1964PhRv..136.1224P}
\begin{equation}
   \label{eq:mergertime}
    t_m=\frac{5D^4c^5}{256 G^3M M_\star (M+M_\star)},
\end{equation}
where $c$ is the speed of light. Therefore, the merger time when $D=D_r$  is
\begin{equation}
\label{eq:resonancetime}
    t_r=\frac{5c^5 (M+M_\star)^{\frac{1}{3}} m^{\frac{8}{3}}}{256 G^{\frac{5}{3}}M M_\star \omega_n^\frac{8}{3}}.
\end{equation}
This expression for the mode resonance time is general and does not depend on which mode becomes resonant with the orbit.
The time before merger when the crust-ocean $i$-mode resonance occurs can be computed simply by substituting eq. \ref{eq:frequencymod} for $\omega_n$.
\subsubsection{Tidal Energy}
When resonance occurs, the amplitude of the oscillation should be maximized. This is directly related to the amount of energy deposited into the ocean due to the tidal force. We can estimate the amplitude at the resonance time by assuming a solution for $a(t)$ of the form $a(t)=G M W_{l m}\frac{Q_{nl}}{A_n^2}c(t)e^{-is\omega_n t}$ \citep{1994MNRAS.270..611L}, where $c(t)$ is a complex valued function of a real variable and $s=\pm1$.
 In terms of $c(t)$, equation \ref{eq:ODEa} becomes \citep{1994MNRAS.270..611L}
\begin{equation}
\label{eq:bfunction}
    \ddot{c}-2 i s \omega_n \dot{c}=D(t)^{-(l+1)}\exp{[i(s\omega t -m \Phi(t))]}.
\end{equation}
Near resonance, numerical solutions have shown that the amplitude increases approximately linearly with time \citep{1994MNRAS.270..611L}. Therefore, neglecting $ \ddot{c}$ and integrating with time gives an approximate expression for $c(t)$
\begin{equation}
\label{eq:cint}
    c(t)\approx\frac{1}{2i s \omega_n}\int D(t)^{-(l+1)}\exp{[i(s\omega t -m \Phi(t))]} dt.
\end{equation}
Assuming $\omega_n>>1/t_r$ (which should be the case as $t_r\gtrsim1$ s and $\omega\gtrsim1$ Hz for reasonable parameters \citep{10.1093/mnras/stad389}), the limits on this integral may be taken as infinite. In this case, the stationary phase approximation may be used to evaluate c(t) \citep{1994MNRAS.270..611L}. The maximum value of c(t) will be 
\begin{equation}
    |c(t)|_{max}\simeq\frac{1}{2\omega_n D_r^{l+1}}\sqrt{\frac{2\pi}{m\ddot{\Phi}(t_r)}},
\end{equation}
where we have evaluated the absolute value of eq. \ref{eq:cint}, and $\ddot{\Phi}(t_r)$ is evaluated at the resonance time. $\ddot{\Phi}$ at time of resonance is
\begin{equation}
    \ddot{\Phi}=\frac{3}{2}\sqrt{\frac{G(M+M_\star)}{D_r^3}}\frac{\dot{D}_r}{D_r}=\frac{3}{8m} \frac{\omega_n}{t_r}.
\end{equation}
This allows us to write $|c(t)|_{max}$ in terms of the parameters of the mode resonance
\begin{equation}
    |c(t)|_{max}\simeq\frac{2}{ D_r^{l+1}}\sqrt{\frac{\pi t_r}{3\omega_n^3}}.
\end{equation}
After tidal resonance in binary inspirals, the energy of the mode should be that of a harmonic oscillator with frequency $\omega_n$ and amplitude $|a(t)|_{max} A_n$. Additionally, for the $l=2$ mode, both the $m=2$ and $m=-2$ modes contribute to the energy equally. Therefore, the tidal interaction will deposit the energy
\begin{equation}
    E=\omega_n^2|a(t)|_{max}^2 A_n^2= \omega_n^2 G^2M^2 W_{lm}^2 \frac{Q^2_{nl}}{A^2_n}|c(t)|_{max}^2
\end{equation}
into the mode \citep{1994MNRAS.270..611L, 10.1093/mnras/stad389}.
The normalization $A_n^2$ and the $l=2$ overlap integral of crust-ocean $i$-modes grow proportionately with the square of the stellar radius and the ocean depth, respectively \citep{10.1093/mnras/stad389}.  Their ratio respects $Q/A_n^2\sim h_o/R_\star$ \citep[e.g.]{2021MNRAS.504.1273P, 10.1093/mnras/stad389}. Hence, the normalization factor can be estimated as 
\begin{equation}
    A_n^2=M_\star R_\star^2,
\end{equation} while the $l=2$ overlap integral is
\begin{equation}
    Q\approx \frac{11}{10}M_\star R_\star^2 \left(\frac{h_o}{R_\star}\right),
\end{equation}
where we infer the prefactor from the results in table 1 of \cite{10.1093/mnras/stad389}. The exact value of the numerical factor is model dependent, but should remain order unity.
The energy in terms of stellar and mode parameters is
\begin{equation}
    E\simeq\frac{121\pi^2}{6400\times2^{\frac{1}{3}}}\frac{c^5 M h_o^2\omega^{\frac{1}{3}}}{(G(M_\star+M))^\frac{5}{3}},
    \label{eq:modeenergy}
\end{equation}
where $\omega_n$ is given by eq. \ref{eq:frequencymod}.
Like $t_r$, the energy deposited into the mode by the tide directly depends on the masses of the objects as well as the depth of the neutron star ocean.

\section{Tidal Resonance as a source of GRB precursor flares}
\label{sec:ModelSummary}
The model we outline in Sec. \ref{sec:tidalresNSO} explains how energy can be deposited into a neutron star ocean through the tidal interaction in the moments leading up to a compact binary merger. \cite{10.1093/mnras/stad389} found that if the energy of the $i$-mode tidal resonance could be released electromagnetically, a  detectable precursor could result. We now extend this picture and apply it to sGRB precursor events. We therefore suppose that gamma-ray precursors exhibiting QPOs could result from this $i-$mode tidal resonance in a neutron star ocean. 

\subsection{Model Parameters}
As we have shown, the ocean tidal resonance is principally described by three quantities: the energy deposited into the mode $E_{tot}$ given by eq. \ref{eq:modeenergy}, the time of resonance $t_r$ given by eq. \ref{eq:resonancetime}, and the mode frequency $\omega_n$ given by eq. \ref{eq:frequencymod}. In a sGRB precursor, these quantities correspond to emission properties. We propose that the energy of the precursor corresponds to the energy deposited into the mode, the time of ignition of the flare corresponds to the resonance time of the $i$-mode, and the QPO frequency is the $i$-mode frequency. {In reality, the precursor energy is a lower limit on the actual energy deposited into the mode, as the radiation efficiency of the emission mechanism remains unknown}. Nevertheless, the precursor energy usefully constrains the total energy deposited into the mode. Using these three observables, we may estimate the parameters of the astrophysical GRB source. 

The three main quantities of our model depend on five system parameters. Four of these parameters directly relate to the neutron star in the binary, while the remaining relates to the companion. Our model is sensitive to the neutron star mass $M_\star$, radius $R_\star$, ocean depth $h_o$, and crust shear modulus to ocean floor pressure ratio $\breve{\mu}/p_o$, as well as the mass of the companion $M$. Excitingly, from the precursor alone, we may constrain parameters essential to understanding the dynamics of the compact binary as well as the interior structure of neutron stars.

\subsection{Model Implications}
\label{sec:GRB211211A}
\begin{figure*}
    \centering
    \includegraphics[width=0.99\linewidth]{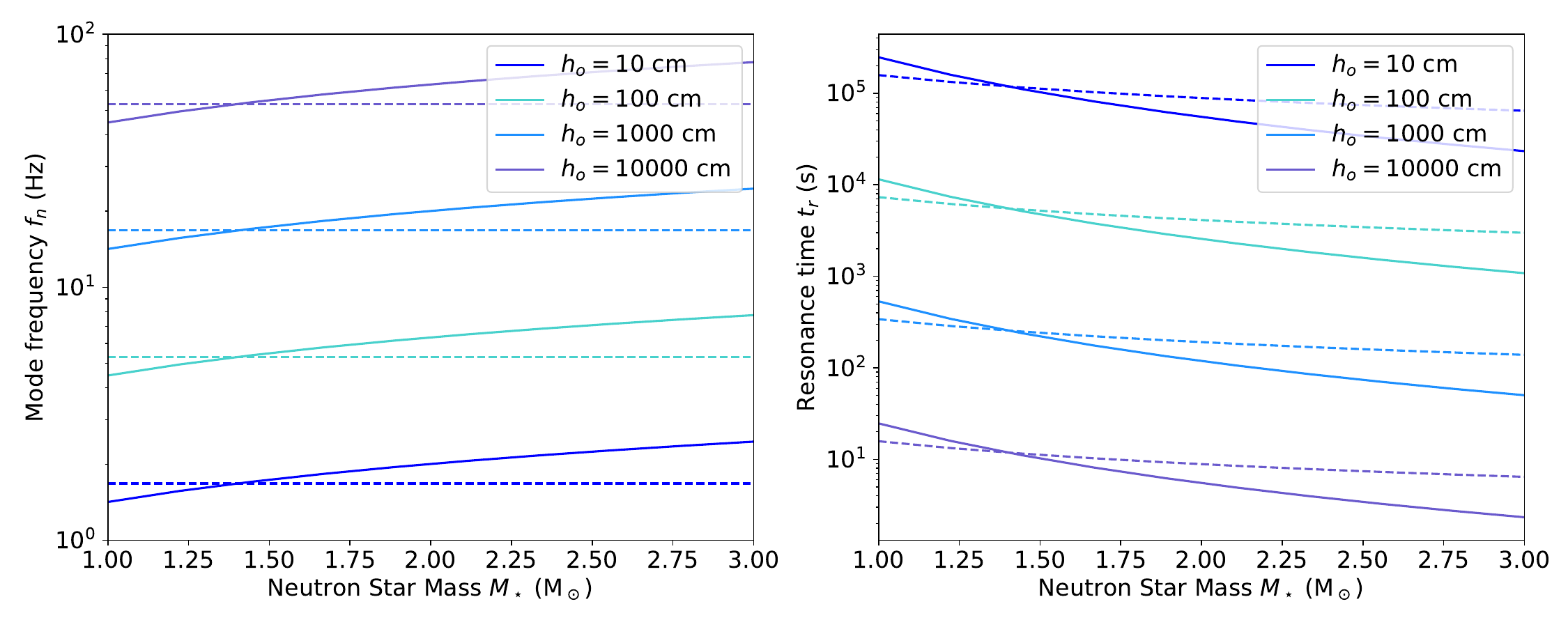}
    \caption{The crust-ocean $i$-mode frequency (left) and resonance time with companion mass $M=1.4$ M$_\odot$ (right) as a function of neutron star mass for the ocean depths $h_o$ shown in the legend. The dashed lines show the mode frequencies and resonance times for the $M=1.4$ M$_\odot$ companion with the same $h_o$. These results assume both neutron stars have $\breve{\mu}/p_o=0.01$ and $R_\star=10$ km.}
    \label{fig:Model}
\end{figure*}
This model for sGRB precursors can describe both binary neutron star and neutron star-black hole mergers. In fact, these two scenarios can be directly distinguished through the companion mass. The only requirement is that one of the component masses in the system be a neutron star, which is already required for sGRBs \citep{2013ApJ...776...18F, 2015IJMPD..2430012R, 2021JPlPh..87a8402A}.

The properties of neutron star oceans have principally been probed by observations of X-ray bursts on accreting neutron stars \citep{1995ApJ...449..800B, 2014ApJ...793L..38S, 2020MNRAS.491.6032C}. The ocean forms at temperatures and densities when the crust melts. To have sizable oceans, neutron star crusts much achieve temperatures of $T\gtrsim10^7$ K, hotter than expected for old neutron stars in compact binaries. The tides can heat neutron stars to temperatures $\sim10^8$ K during the inspiral \citep{1994MNRAS.270..611L}, and potentially higher depending on the viscosity in the neutron star \citep{1992ApJ...397..570M}. 

With this model, we directly probe the depth of the ocean $h_o$ via compact binary coalescence. $h_o$ is very sensitive to the material that composes the neutron star crust, the neutron star crust temperature $T$, as well as the equation of state at the neutron star surface \citep{1993PhRvE..47.4330F, 1995ApJ...449..800B, 2007ASSL..326.....H,
2009PhRvL.102s1102H, 2018MNRAS.480.5511B, 2020PhRvD.101j3025G}  (although there is degeneracy between these three quantities). Constraints on the ocean depth for neutron stars in compact binaries can inform whether there is a difference in ocean structure between neutron stars in X-ray binaries and in compact binaries. 

While this model is sensitive to five extremely interesting properties of neutron stars and compact binaries, its reliance on only three main observables limits its parameter estimation ability. For certain reasonable choices of neutron star mass and radius $M_\star$ and $R_\star$,  one can solve for the other three parameters with this model, and immediately distinguish an NSBH from a BNS based on the companion mass results. An alternative approach might be to solve for the neutron star mass and radius as well as the companion mass as a function of $h_o$ and $\breve{\mu}/p_o$. Estimates of the neutron star mass $M_\star$ and $R_\star$ are particularly exciting as they can directly be used to constrain the neutron star equation of state.

The degeneracy in parameters can nevertheless be broken in multiple ways. Most promising is a coincident GW detection from the merger. Chirp mass and total mass measurements, as well as tidal deformability limits from GWs \citep[e.g.]{2019PhRvX...9a1001A} provide additional constraints on the system which can completely disentangle all parameters. In the case of a binary neutron star merger, the oceans of the two different neutron stars may become resonant at different times, causing two precursors with QPOs. {In Fig. \ref{fig:Model}, we show the mode frequency $f_n=\omega_n/2\pi$ and resonance time $t_r$ as a function of neutron star mass for different values of $h_o$ with a $1.4$ M$_\odot$ companion. $f_n$ and $t_r$ are also shown for the companion with $h_o$ values. Both $f_n$ and $t_r$ can differ by at least order unity between the two stars, so the two precursors can be distinguished if their durations are $\lesssim10 \%$ of $t_r$ \citep{2010ApJ...723.1711T}.} This then provides six equations to disentangle eight parameters. 
Most interestingly, if the degeneracy between $M_\star$ and $R_\star$ can be broken with a GW detection, our model constrains the equation of state \citep{2001ApJ...550..426L, 2018PhRvL.121p1101A, 2021ARNPS..71..433L} by providing more data points to directly probe the neutron star mass-radius relationship. 

The precursors should be distinguishable from the main emission for sources of interest. We take an NSBH merger with a 1.4 M$_\odot$ neutron star and  a 5 M$_\odot$ black hole, which should be a sGRB progenitor \citep{2011ApJ...727...95P}, as an example. The depth of a relativistic degenerate neutron star ocean is \citep{1995ApJ...449..800B}
\begin{equation}
\label{eq:oceandepth}
    h_o\approx 12.8 \text{ m} \left(\frac{A}{12}\right)^{-1}\left(\frac{Z}{6}\right)^{-\frac{1}{3}}\left(\frac{T}{10^7 \text{ K}}\right),
\end{equation}
where $Z$ and $A$ are the atomic number and mass of the ions that compose the crust, respectively. For fiducial ocean values of $T=10^7$ K, $A=12$, and $Z=6$ as well as neutron star properties $R_\star=10$ km and $\tilde{\mu}/p_o=0.01$, the precursor parameters are $f_n=\omega_n/2\pi=20$ Hz, $t_r=70$ s, and $E_{tot}=4\times10^{47}$ erg. For a BNS where the companion is $M=1.4$ M$_\odot$, the energy of the precursor would remain approximately unchanged while the time before merger would increase to 3 min, allowing the scenarios to be distinguished. {The value inferred for the} companion mass $M$ is most sensitive to $t_r$ while the inferred $h_o$ value is sensitive to $E_{tot}$ and $\omega_n$. The other neutron star parameters may be constrained by $\omega_n$.

\subsection{Electromagnetic Emission Mechanisms}
In its current form, our model remains agnostic to how precursor gamma-rays are generated. Our model also does not predict that the expected emission will necessarily be in the form of gamma-rays.  We choose gamma-rays as the application of this model because Fermi-GBM's all-sky field of view makes the instrument particularly well-suited to detecting rapid transients \citep{2009ApJ...702..791M}. In principle, the resultant electromagnetic emission from a neutron star ocean tidal resonance could be across the electromagnetic spectrum. The connection between our model and electromagnetic emission remains speculative at this stage.

To actually ignite an electromagnetic flare, we envision a scenario in which the energy deposited into the neutron star ocean by the tide excites particles on the neutron star surface to high energies. The resultant high-energy electrons on the surface may synchrotron radiate in the presence of the strong surface magnetic field. {The ocean Alfv\'en frequency $\omega_A\sim l (l+1)B^2/4\pi \rho_o R^2$ may be comparable to the crust-ocean $i$-mode frequency for $B\sim10^{12}$ G. Consequently, high magnetic field can modify the surface structure and $i$-mode properties as well as complicate connecting tidal energy deposition with emission.} 

The energy deposited by the tide may nevertheless be comparable to the breaking energy of neutron star crusts, which ranges from $10^{44}-10^{46}$ erg \citep{2012PhRvL.108a1102T, 2018MNRAS.480.5511B}, causing the crust to crack or melt \citep{2012ApJ...749L..36P}. {While full crustal failure may be difficult to achieve initially since only $0.1\%$ of the crust-ocean $i-$mode energy is deposited into the crust \citep{2005ApJ...619.1054P},} the back reaction of the strongly deformed or even damaged crust may cause the crust-core $i$-mode frequency to increase as the mode penetrates deeper into the star like the $r$-mode under the influence of a strong magnetic field \citep{2000ApJ...534L..75A, 2001PhRvD..64j4013R, 2001PhRvD..64j4014R}. This can allow for the extraction of more tidal energy as the overlap integral grows. If the neutron star crust breaks, subsequent reconnection of the liberated crustal magnetic fields may induce large scale particle acceleration and consequently the emission of gamma-rays \citep[e.g.]{2015MNRAS.449.2047L, 2017ARA&A..55..261K}. {Each of these mechanisms likely has a different radiation efficiency which can affect our energy estimate.} We leave the details {including the effects of strong magnetic fields} for future work. Whatever the mechanism, the energetics of the resonant tide \citep{10.1093/mnras/stad389} coupled to the exotic conditions on neutron star surfaces make electromagnetic emission a plausible result of ocean-tidal resonances.

\section{Application to GRB 211211A}
\label{sec:GRB211211A}
\begin{figure*}
    \centering
    \includegraphics[width=0.99\linewidth]{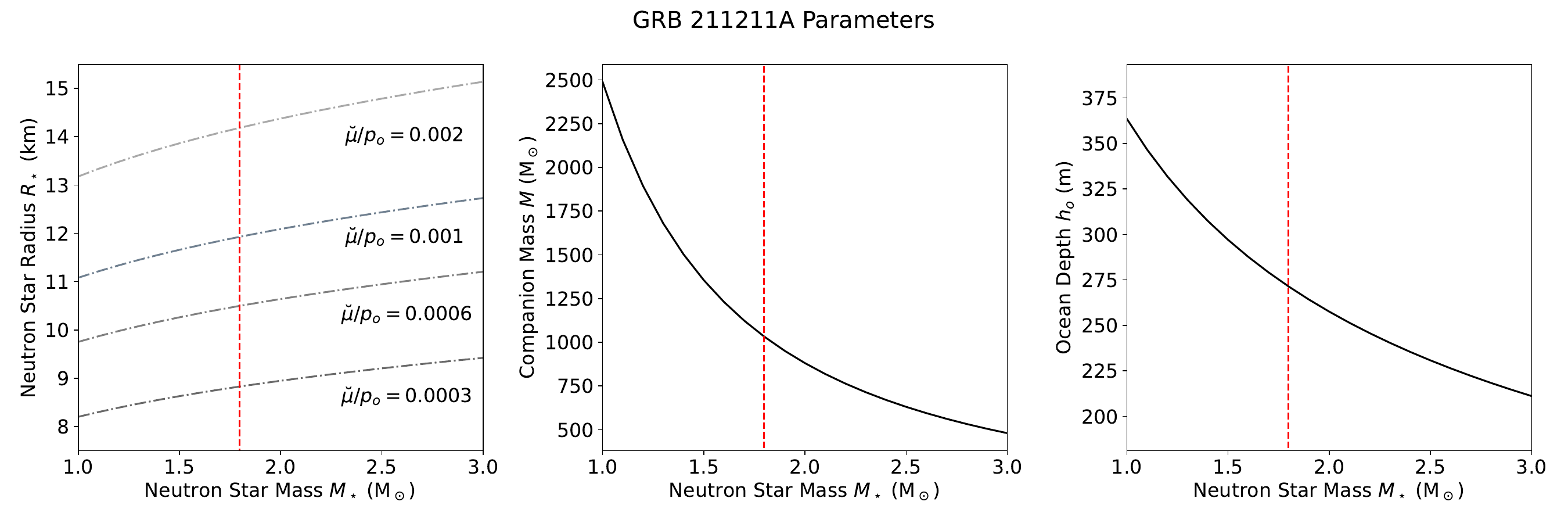}
    \caption{The parameters of the compact binary coalescence which produced GRB 211211A estimated by the precursor model presented in this paper. For the range of plausible neutron star masses, 1-3 M$_\odot$, we determine the mass of the companion $M$ (middle), the depth of the neutron star ocean $h_o$ (right), and the radius of the companion neutron star $R_\star$ (left). Determining $R_\star$ also requires choices of $\breve{\mu}/p_o$, the ratio of the neutron star crust shear modulus to the pressure at the crust-ocean boundary. The red dashed line on each panel corresponds to the neutron star mass $M_\star=1.8$ M$_\odot$, below which the Schwarzschild radius of the companion exceeds the resonance binary separation.}
    \label{fig:Params}
\end{figure*}
GRB 211211A is one of the longest sGRB events detected with a burst duration of 51.37 s \citep{2022Natur.612..223R}. The subsequent detection of an associated kilonova  located at distance of 350 Mpc suggests that the cause of this burst was a compact binary coalescence \citep{2022Natur.612..223R, 2022Natur.612..228T, 2022Natur.612..232Y, 2022Natur.612..236M, 2022ApJ...933L..22Z}. A precursor flare 0.9 seconds prior to the initiation of GRB 211211A was also detected, {possibly} exhibiting QPOs with frequency 22 Hz \citep{2022arXiv220502186X}. The precursor flare had an isotropic equivalent energy of $7.7\times10^{48}$ erg while the isotropic equivalent energy of the main event and extended emission approached $10^{52}$ erg \citep{2022arXiv220502186X}.

Many questions about this system remain, as no model has conclusively determined the source of this event. Suggested sources of GRB 211211A include a NSBH merger \citep{2022ApJ...934L..12G, 2022ApJ...936L..10Z, 2023arXiv230614947G}, a BNS merger involving a magnetar \citep{2022arXiv220502186X, 2022ApJ...939L..25Z}, a white dwarf-neutron star merger \citep{2023ApJ...947L..21Z} and a collapsar \citep{2023ApJ...947...55B}. BNS magnetar models invoke a shattering flare induced by the tides of the companion and cracking the magnetar crust \citep{2022A&A...664A.177S}, while NSBH models invoke both the presence of a magnetar \citep{2022ApJ...934L..12G} and the presence of a rapidly spinning BH \citep{2022ApJ...939L..25Z} to enhance energy release from tides.
\cite{2023arXiv230510682Z} investigated the plausibility that the source of GRB 211211A was a strangeon star and invoked tidally induced crust fracturing to explain the energetics of the system. These models broadly rely on large tidal forces and strong magnetization to explain both the precursor and extended length of the main emission. As more events like this one are observed \citep[e.g.]{2023arXiv230702996D}, models will likely converge on an explanation of GRB 211211A and other similar sources \citep{2023arXiv230900038G}.

As {possibly} the first sGRB observed with QPOs during its precursor and without a conclusive description of this system, GRB 211211A represents an excellent test bed for our precursor model. Within the context of our model, the GRB 211211A precursor would be interpreted as a flare induced in a neutron star ocean by resonant tides. The pulsating ocean gives rise to the QPOs in the gamma-ray emission. This picture qualitatively agrees with that of \cite{2022A&A...664A.177S} in which tidal forces crack the neutron star crust and release energy for a flare. \cite{2022A&A...664A.177S} associate the QPOs with resonant torsional $g$-modes of a highly magnetized neutron star surface after crust cracking. High magnetization is required to explain the nonthermal spectrum of the precursor and provide sufficient energy to the flare. \cite{2022A&A...664A.177S} rely on the mass results of the kilonova modeling  of \cite{2022Natur.612..223R} and therefore consider tidal resonances in a BNS only, despite ambiguity in the literature. Our model by contrast leaves open the possibility that the system is a NSBH and assumes tidal resonance of the crust-ocean $i$-mode. 

The observations of GRB 211211A provide a precursor ignition time, oscillation frequency, and total energy \citep{2022arXiv220502186X}, the exact parameters our model requires. The $22$ Hz QPO frequency resembles magnetar crustal shear mode frequencies, and thus represents a plausible value for that of a surface ocean mode \citep{2007MNRAS.374..256S, 2011MNRAS.414.3014C}. {While the QPO remains unconfirmed \citep{2023arXiv231012875C},} we apply our model to this system {as an example of how it can be used} and check whether it reasonably explains phenomena of this sGRB. Without any GW emission to unambiguously measure the source masses, we cannot constrain all system parameters. Note, however, that $t_r$ (eq. \ref{eq:resonancetime}) is only a function of the two companion masses. Consequently, to obtain $M$, one only needs $M_\star$. With $M$ and $M_\star$, $h_o$ can immediately be solved for from the observed $E$. Therefore, setting $E=7.7\times10^{48}$ erg, $t_r=0.93$ s, and $f_n=\omega_n/2\pi=22$ Hz, we solve eqs.  \ref{eq:frequencymod}, \ref{eq:resonancetime}, and
\ref{eq:modeenergy} for $R_\star$, $M$, and $h_o$ for $M_\star\in [1 $ M$_\odot,$ 3 M$_\odot]$—viable masses for a neutron star—and choices of $\breve{\mu}/p_o$. We neglect the effects of gravitational and cosmic redshift on $t_r$ and $f_n$. This affects our results by no more than $30\%$, exceeding the precision needed to assess our model's implications. We show our parameter results in Fig. \ref{fig:Params} as a function of $M_\star$.

The main prediction of our model is that the source of GRB 211211A is an extreme-mass ratio NSBH merger. This prediction comes directly from associating the precursor to a tidal resonance, particularly the resonance of a mode with the {alleged} QPO frequency, and is independent of the exact nature of the excited mode. Therefore, if the resonant neutron star mode which ignites the precursor has frequency $\sim20$ Hz (irrespective of which mode and how), the companion mass must exceed $\sim 500$ M$_\odot$ simply by eq. \ref{eq:resonancetime}. For an NSBH, resonance must occur before the neutron star crosses the horizon of the companion black hole. This constrains the parameter space to companions with mass below $1000$ M$_\odot$ and consequently neutron stars with mass $M_\star\gtrsim1.8$ M$_\odot$. The viable regions of parameter space are to the right of the red-dashed line in Fig. \ref{fig:Params}. Again, this constraint is independent of the nature of the mode, and only relies on associating the QPO with the resonant mode which causes the precursor.

Associating the precursor with the crust-ocean $i$-mode subsequently informs the neutron star structure. Our model predicts a neutron star ocean with $h_o\gtrsim200$ m. Such a deep ocean ensures that the tidal overlap integral of the crust-ocean mode is large enough to garner sufficient energy for the precursor. This particularly deep ocean suggests that the temperature inside the neutron star should be very high: $T\gtrsim 2\times10^8$ K for a crust made of carbon and hotter for heavier elements (see eq. \ref{eq:oceandepth}). This is comparable to surface temperatures reached during accretion \citep{1984ApJ...278..813F, 1990A&A...227..431H, 2003A&A...404L..33H, 2008A&A...480..459H}, and would require extreme heating. {Tidal heating by core $g-$mode resonances can produce these temperatures \citep{1994MNRAS.270..611L}; however, the core $g-$mode frequencies are likely $>20$ Hz and would be resonant after the crust-ocean $i-$mode. Alternatively, accretion onto the neutron star, if the binary is in a gaseous environment such as an active galactic nucleus (AGN) accretion disk, may heat the surface.} 

The mode frequency gives the neutron star radius with certain choices of $\breve{\mu}/p_o$. We see that radii consistent with plausible neutron star equations of state \citep{2020Sci...370.1450D} must have $\breve{\mu}/p_o\approx0.0005-0.002$. These values imply a particularly weak crust compared to the internal pressure of the neutron star, and are broadly consistent with a deeper ocean whose ocean floor pressure is greater. The weak shear modulus implies that the mode penetrates deeper into the crust, which could account for the large value of $h_o$ needed to provide the flare energy. A lower value of $\breve{\mu}$ also decreases the energy needed to fracture the crust \citep{2018MNRAS.480.5511B}, making a shattering event much more likely. Such a large amount of energy tidally deposited into the neutron star surface, along with a low breaking energy, leaves a majority of the energy for the precursor, again showing that the model is self-consistent.  

If our model accurately describes the event, GRB 211211A would represent the first ever detection of an extreme mass ratio compact binary inspiral. Our model also suggests a high mass neutron star in the event, which has particular relevance to constraining the neutron star equation of state \citep[e.g.]{2007PhR...442..109L, 2013ApJ...765L...5S, 2023arXiv230606218B}. For such a heavy black hole and a neutron star to form in binary, a dynamical environment such as an AGN or globular cluster would likely host the source \citep[e.g.]{2021ApJ...920L..42G}. This extremely exotic potential origin could explain the peculiarity of this event. To account for such a small fraction of GRBs, long sGRBs may require extreme tides and unique neutron star parameters. Furthermore, previous work has suggested NSBH origin for this event \citep{2022ApJ...934L..12G,2022ApJ...939L..25Z, 2023arXiv230614947G}, as the distinction between shorter and longer sGRBs may correspond to the difference between BNS and NSBH merger origin \citep{2023arXiv230412358D}. 

Such a large mass ratio inspiral is nevertheless extremely difficult to explain given the observed sGRB and kilonova. These transients require full tidal disruption of the neutron star, which intermediate mass black holes should fail to cause \citep{2022MNRAS.514.5385N}. In fact, $M\sim10$ M$_\odot$ represents an upper limit on the companion mass which can plausibly cause a sGRB \citep{2011ApJ...727...95P, 2022MNRAS.514.5385N}. Furthermore sGRBs from NSBH mergers should be rare, with event rates only as high as $\sim100$ Gpc$^{-3}$ yr$^{-1}$ \citep{2023PhRvX..13a1048A, 2023arXiv230614974B}. Future numerical simulations of extreme mass ratio inspirals \citep[e.g.]{2023PhRvD.108h4015B} of NSBHs will hopefully provide further insight into the feasibility of the scenario we consider.

If tidal resonance actually ignites the precursor, a higher frequency resonant mode would provide a more plausible companion mass estimate. For example, a resonant mode with frequency $\sim50$ Hz \citep[e.g.]{2022A&A...664A.177S} would imply a companion mass $\sim10$ M$_\odot$, which could plausibly produce the sGRB. This leaves open the possibility that the 22 Hz QPOs represent the decaying after shocks of a previously excited mode. The QPO may represent a previously excited crust-ocean $l=0$ mode, so that the $l=2$ crust-ocean $i$-mode has frequency $\sim50$ Hz. This would predict an ocean depth of $h_o\sim60$ m and a more modest crust temperature $T<10^8$ K. 
Alternatively, the QPO could correspond to a much earlier tidally excited $l=2$ crust-ocean $i$-mode, while a resonant crustal shear mode or the core-crust $i$-mode shatters the crust \citep{2012PhRvL.108a1102T}. 
If this is the case, fainter precursors may have preceded the observed event by $\gtrsim1$ min \citep{10.1093/mnras/stad389}. Searches for earlier faint emission among Fermi-GBM, \textit{Swift}, and Insight-HXMT sub-threshold data may reveal further evidence of our model at work in this system.

\section{Conclusion}
\label{sec:Conclusion}
We have presented a new model for sGRB precursors invoking tidal resonance of the surface mode of a neutron star in a compact binary coalescence. Our  model posits that precursors to sGRBs can be ignited by the resonance of the neutron star crust-ocean $i$-mode with the orbitally modulated tidal forces from the inspiraling companion. In this picture, the energy fueling the flare is deposited by the tide and QPOs emerge as a natural consequence of the excitation of the mode. Thus our model can be applied to any precursor with QPOs.

With three main observables, our model can provide constraints on compact binary parameters solely from information about the precursor. Companion mass constraints and by extension the type of merger in question can be obtained from just the observed time prior to the main sGRB and the frequency of the QPOs, which corresponds to the resonance frequency of the mode in our model. By associating the precursor with the crust-ocean $i-$mode specifically, we constrain the neutron star ocean depth, neutron star radius, and even shear modulus of the neutron star crust.

Our model provides an interesting, though likely inaccurate, explanation for some of the observable properties of GRB 211211A.  If true, GRB 211211A would be associated with an NSBH merger with an intermediate mass black hole and a high mass neutron star. Such a system would be the first of its kind, representing the discovery of an intermediate mass black hole as well as the largest black hole involved in a compact binary merger (excluding those potentially observed by pulsar timing arrays \citep{2023arXiv230616222A}). For the excited $i-$mode to contain the energy, a deep neutron star ocean would be needed, suggesting that the crust must either be composed of lighter elements than previously considered or extremely hot \citep[e.g.]{2008LRR....11...10C}. Such a deep ocean also necessitates a small shear modulus for reasonable neutron star radii. Some distinctive features our model identifies may help explain the event's extraordinary status as a very long sGRB. 

The predictions of our model nevertheless remain difficult to reconcile with the sGRB main emission and the observed kilonova. As we have discussed, such a large mass companion would have difficulty tidally disrupting the neutron star to produce powerful electromagnetic emission during the GW-driven merger. We have found that the {claimed} QPOs are extremely unlikely to be the mode that caused the precursor, although it remains possible that a different resonant mode did. {Because the emission mechanism is unknown, however, it remains just as likely that the QPOs originate from intrinsic GRB properties rather than pulsational modes.}
Unfortunately, the lack of observed GW emission keeps the origins of GRB 211211A nebulous \citep{2023MNRAS.518.5483S}. The joint detection of GWs with this sGRB could have more definitively constrained the applicability of our model as well as other proposed models to this unique system.

Applying our model to more sGRBs precursors, especially those with long durations and other distinctive features \citep{2023ApJ...954L...5V}, may yield interesting results. As detection techniques improve, more sGRBs will be identified \citep{2023arXiv230701103K}, hopefully providing more opportunities to test our model.  
Previous searches for QPOs in sGRB precursors have yet to reveal any additional candidates with $>3\sigma$ significance \citep{2022ApJ...941..166X}, but have been constrained by photon statistics. Continued observations, particularly in coincidence with detected GW events during the O4 run of LIGO-Virgo-Kagra \citep{2022ApJ...937...79C}, as well as improved targeted searches will hopefully reveal more such candidates. If more sGRB precursor QPOs can be identified, models of tidal resonance-induced precursor emission like the one presented in this paper can immediately be tested.

\section*{Acknowledgments}

The authors are grateful to Nils Andersson, Roger Blandford, and Roger Romani for reading and providing feedback on this manuscript. The authors thank Isabella Leite for helpful discussions and reviewing the manuscript. The authors thank Stanford University, Columbia University in the City of New York, and the University of Florida for their generous support. The Columbia Experimental Gravity group is grateful for the generous support of Columbia University. AS is grateful for the support of the Stanford University Physics Department Fellowship and the National Science Foundation Graduate Research Fellowship Program. LMBA is grateful for the Columbia Undergraduate Scholars Program Summer Enhancement Fellowship and the Columbia Center for Career Education Summer Funding Program.

\section*{Data Availability}
The data underlying this article will be shared on reasonable request to the corresponding author.

\bibliography{Refs} %
\label{lastpage}
\end{document}